\documentclass[conference,a4paper,twocolumn]{IEEEtran}
\def\IEEEkeywordsname{Keywords }
\ifCLASSINFOpdf
\else
\fi
\usepackage{algorithm,algpseudocode,algorithmicx,tabularx}
\usepackage{setspace}
\usepackage{graphicx}
\usepackage{amssymb}
\usepackage{amsmath}
\usepackage[justification=centering]{caption}

  \usepackage{cleveref}

\crefformat{section}{\S#2#1#3} 
\crefformat{subsection}{\S#2#1#3}
\crefformat{subsubsection}{\S#2#1#3}
\makeatletter
\newcommand{\multiline}[1]{%
  \begin{tabularx}{\dimexpr\linewidth-\ALG@thistlm}[t]{@{}X@{}}
    #1
  \end{tabularx}
}
\begin{document}
\title{Optimized Blockchain Model for Internet of Things based Healthcare Applications}

\author{
\IEEEauthorblockN{
Ashutosh Dhar Dwivedi$^{{1},{2}}$, Lukas Malina$^{3}$, Petr Dzurenda$^{3}$ and Gautam Srivastava$^{2,4}$
}\\

\IEEEauthorblockA{
$^{1}$Institute of Computer Science, Polish Academy of Sciences, Warsaw, Poland\\
$^{{2}}$Department of Mathematics and Computer Science, Brandon University, Brandon, Canada\\
$^{3}$Department of Telecommunications, Brno University of Technology, Brno, Czech Republic\\
$^{{4}}$Research Center for Interneural Computing, China Medical University, Taichung, Taiwan, Republic of China\\
}
\thanks{Research described in this paper
was financed by the National Sustainability Program under grant LO1401
and Ministry of Interior under grant VI20172019093.}}
\markboth{}{}
\pagestyle{empty}%
\maketitle%
\thispagestyle{empty}

\begin{abstract}
There continues to be a recent push to taking the cryptocurrency based ledger system known as Blockchain and applying its techniques to non-financial applications. One of the main areas for application remains Internet of Things (IoT) as we see many areas of improvement as we move into an age of smart cities. In this paper, we examine an initial look at applying the key aspects of Blockchain to a health application network where patients health data can be used to create alerts important to authenticated healthcare providers in a secure and private manner. This paper also presents the benefits and also practical obstacles of the blockchain-based security approaches in IoT. 
\end{abstract}


\IEEEpeerreviewmaketitle

\section{Introduction}
We find ourselves now in a new age of technology. After the mobile Internet technologies and the World Wide Web owned much of the last $20$ years, the time has come for the Internet of Things (IoT) revolution. IoT describes a network where every object is uniquely identified, accessible and becomes part of the Internet. IoT consists of devices responsible for generating, processing and exchanging privacy-sensitive information. It has a broad range of applications including health management, smart homes, traffic, agriculture, and weather monitoring to name just a few. A vast variety of smart IoT devices are used on each layer of Internet Technology. These IoT devices collect and analyze a lot of sensitive data and therefore security and privacy of information is one of the essential requirements. IoT devices are lightweight and also have shallow energy footprints. This small amount of available energy is generally used to execute core application functionality and therefore supporting other challenges like security and privacy is quite challenging. Due to the decentralized topology and resource constraints of devices, conventional security and privacy approaches are inapplicable.
Consequently, IoT demands scalable, lightweight and distributed privacy and security safeguard. The blockchain technology that underpins Bitcoin \cite{nakamoto2012bitcoin}, has the potential to overcome the problems mentioned above due to its secure, distributed, immutable, transparent and auditable ledger. The blockchain protocol structures all the information in a chain of connected blocks, where each block can store transactions related to its specific application. Blocks are linked together by a reference to the previous block, forming a chain. Once the block is full of transactions, it is appended to the blockchain through a \emph{mining} process. The mining process is performed by some specific nodes known as \emph{miners}, by solving a resource consuming mathematical puzzle called \texttt{Proof of Work} (PoW). However, applying the blockchain in the context of IoT is not straightforward. It has several challenges which must be overcome such as low scalability, high resource demand, and traffic overhead just to name a few. 

In this paper, we introduce a novel blockchain model that is optimized for IoT devices. To exemplify our idea, we use the scenario of Remote Patient Monitoring (RPM). RPM provides a healthcare facility access to the patient outside the conventional clinical setting (in the home as an example). A patient is equipped with wearable IoT devices, and these devices can provide information to healthcare providers such as blood glucose level, blood pressure, breathing pattern and many others. In our model, we eliminate the issues of blockchain technology and try to install a lightweight blockchain that retains the underlying privacy and security benefits. In our naive model, we eliminate the concept of \texttt{Proof of Work} (PoW). Therefore, security and privacy of the proposed framework relies on the distributed property of the model. Our model consists of five key parts: the Blockchain Network, Cloud Storage, Healthcare Providers, Smart Contracts and Patients equipped with healthcare wearable IoT devices. To decrease network overhead and delay we divide our blockchain into clusters through an overlay network. Instead of using a single blockchain, we use clusters, and each cluster is a group of several nodes having one node treated as a Cluster Head. 

The rest of the paper is organized as follows. In \cref{rel}, we discuss some key related works in the context of our work here. We then analyze blockchain based security approaches in \cref{ana}. Our main results are presented next in \cref{res}. We follow these results with a discussion around which transaction types will be allowed on our model in \cref{trans}. Last, we end with some concluding remarks in \cref{conc}

\section{Related Work}\label{rel}
The investigation of blockchain technologies in IoT has been conducted in several papers.

Conoscenti \textit{et al.} \cite{conoscenti2016} survey the blockchain and Peer-to-Peer (P2P) approaches and their potential usage in a private-by-design IoT. Their investigation detects several issues in the integrity, anonymity and adaptability of blockchain systems. They conclude that large blockchain systems (e.g. Bitcoin) are secure but not appropriate for IoT due to scalability issues. Furthermore, in \cite{kshetri2017} we see how blockchain can address some general IoT challenges. For example, blockchain and smart contracts can improve security in supply chain networks in IoT. Reyna \textit{et al.} \cite{reyna2018} study IoT and blockchain integration and its challenges. The survey work summarizes and discusses IoT devices that can be used as blockchain components, important blockchain technologies and recent IoT–Blockchain applications. The most IoT-Blockchain applications are based on Ethereum. 

Huh \textit{et al.} \cite{huh2017} propose a novel blockchain PKI management system using Ethereum and smart contracts. The security solution uses RSA public key cryptosystems where public keys are stored in Ethereum and private keys are saved on individual devices. Nevertheless, the presented solution has two problems. The first problem is the long duration time of one transaction (approximately $12$ seconds). The second problem is a large storage requirement for the constrained Hardware of IoT clients. Solving this problem by a proxy node which should be the third trusted party then decrease the level of security.

Dorri \textit{et al.} \cite{dorri2017block} present a lightweight instantiation of a blockchain in a smart home setting. The solution defines local private blockchains that store policy and transaction information in an immutable ledger. The proposed local blockchain does not use PoW and is controlled by the owner. The local blockchain is managed by a local miner in each smart home. These miners have a list of devices and process all transactions to and from smart homes. The unicast communication between devices is secured with a shared key but these keys are also generated by the local miner. The similar general framework based on blockchain for broader IoT applications is proposed in another Dorri \textit{et al.} publication \cite{dorri2017tow}. 

In our paper, we propose a blockchain-based security solution for a health application network where patients share their health data with healthcare providers in a secure and private manner. 

\section{Analysis of Blockchain-based Security Approaches in IoT}\label{ana}
The core properties of blockchain such as a distributed ledger, public key cryptography and consensus algorithm make blockchain promising for IoT~\cite{8624250}. Data transactions take place with multiple networks instead of a centralized body. This is the reason blockchain records are naturally transparent and can be analyzed and tracked by anyone authorized on the network. However, applying blockchain in the context of IoT is not straightforward, and there are a few challenges with this as well. We discuss both aspects for applying blockchain to IoT in this section.  
   
\subsection{Positives and Security Benefits}
\begin{itemize}
  \item \textbf{Privacy/Anonymity}- Blockchain uses the digital identity of the transactions using public key cryptography. This mechanism hides the real identity of IoT applications with sensitive data. 
  
  \item \textbf{Trustworthiness} - The data of IoT applications are transported through infrastructure owned by multiple organizations~\cite{s19020326}. This is required to monitor the IoT applications data to improve the services provided by organizations. Traditionally supply chain monitoring rely on centralized architecture but blockchain's decentralized ledger provide more trust while moving assets (real or digital) through infrastructure owned by multiple and diverse stakeholders.    
  
  \item \textbf{Smart contracts} - Some blockchain networks provide ``smart contract" facilities, such as Ethereum, allowing the creation of agreements which will be executed when conditions are met. For example, one system is authorized to pay if certain provided conditions are met (services delivered). These smart contracts are embedded in the system. 
	 \item \textbf{DDoS notification and mitigation} - Smart contracts and blockchain can be used in a collaborative architecture which provides \texttt{DDoS} notification across multiple domains. For example, the work in~\cite{rodrigues2017blockchain} proposes a solution based on smart contracts that advertise white or blacklisted \texttt{IP} addresses between users and autonomous
systems in a public and distributed infrastructure using software-defined network technology. Furthermore, trusted blockchain-based transactions prevent attackers to directly install malware on IoT devices and establish their IoT botnets in order to launch massive \texttt{DDoS} attacks. In addition, checking outgoing traffic prevents the spread of \texttt{DDoS} messages from IoT devices.
\end{itemize}

\subsection{Negatives and Practical Obstacles}
\begin{itemize}
\item \textbf{Resources limits} - IoT devices have limited memory and computational power while blockchain demands excessive resources. The computational requirements for mining blocks in blockchain are well beyond the capabilities of resource-constrained IoT devices. 

\item \textbf{Bandwidth limits} - Due to the decentralized nature of blockchain, nodes in the network exchange information to validate the transactions. IoT devices operating at the end-device layer have limited bandwidth constraints. Some Edge-devices may have enough bandwidth, but the bandwidth requirements of blockchain may exceed those upper thresholds sometimes.    

\item \textbf{Connectivity limits} - In blockchain technology, all devices stay connected with the blockchain network and cooperate through pre-defined protocols. This always-connected feature of blockchain technology makes IoT devices potentially more susceptible to security attacks.  

\item \textbf{Memory limits} - Most open blockchain technologies charge a transaction fee and use it to reward the nodes involved in mining blocks. But in the scenario of healthcare applications, our needs and restrictions are much different. Health data may be measured very frequently. Storing \textbf{ALL} health data for a large number of patients on the chain may prove to be a space sensitive issue.
\end{itemize}

\section{Blockchain-Based Security Solution for Remote Patient Monitoring System}\label{res}

\begin{figure}[h]
    \centering
    \scalebox{.70}{\includegraphics{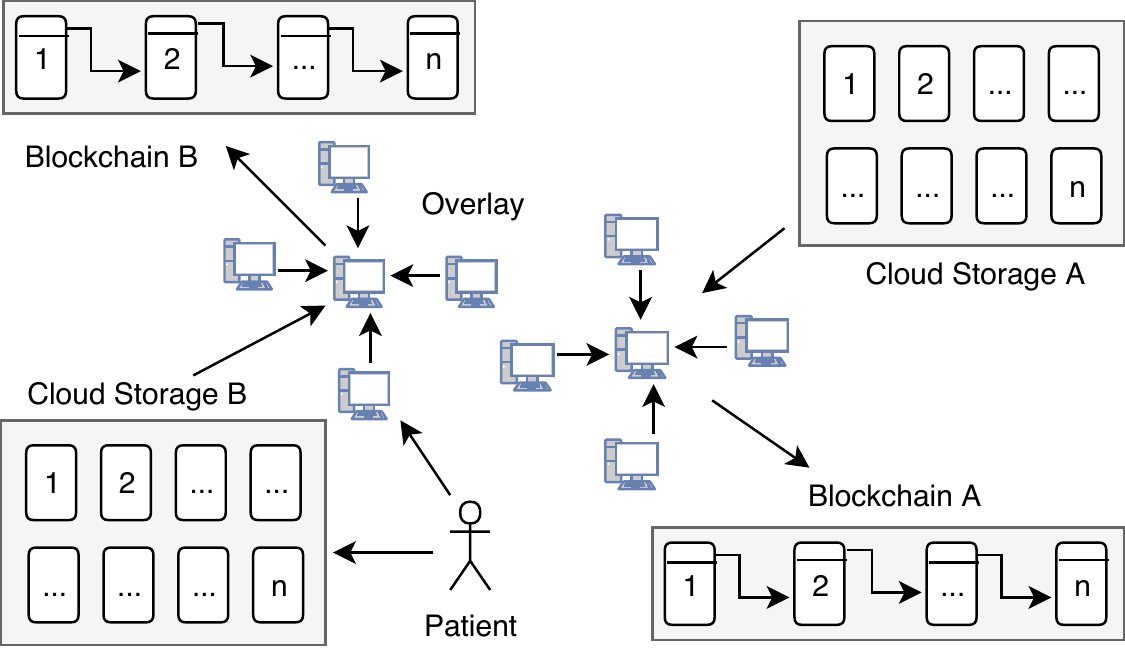}}
    \caption{Overlay Network} 
      \label{overlay}
  \end{figure}

In this section, we present a blockchain-based security solution for concrete IoT systems, namely, a remote patient monitoring system.
We consider a typical remote patient monitoring system where patients are equipped with wearable healthcare IoT devices. These devices will collect health data of patients such as heartbeats, walking distance, or sleeping conditions. The patient is responsible for granting, denying or revoking data access from any other parties such as healthcare providers. If a patient needs treatment, he can share his data with desired healthcare providers. Once the treatment is over, he can remove data access by the network. The proposed architecture, shown in Figure~\ref{overlay}, includes three-tiers, namely the patient (patient equipped with wearable devices), the overlay network, and cloud storage.

\subsection*{Patient wearable devices:} Wearable devices consist of micro-controllers that can be worn on the body or could be embedded into clothing as well. These devices are user friendly and could be connected with wireless data transmission or mobile devices and provide real time feedback and alerting mechanisms. The wearable devices do not always send data to the network itself but can gain assistance of smart contracts to transmit data when necessary.

\subsection{Deployment of Smart Contracts}

Smart contracts are a transaction automation algorithm built on the blockchain. Smart contracts enable transacting parties to set terms when a transaction could be executed automatically. Smart contracts can be deployed in a variety of platforms. In other words, smart contracts  allow  users  to  execute  a  script  on  a  blockchain network  in  a  verifiable  way  and  allows  many  problems  to  be solved in a way that minimizes the need for trust. To do so, they allow users to place trust directly in the deterministic protocols and promises specified in a smart contract, rather than in a third party \cite{hanada2018smart}. A  smart  contract  has  its  own  address  and  account  on  the blockchain. Consequently, it can maintain its own state and take ownership  of  assets  on  the  blockchain,  which  allows  it  to act as an escrow. Smart contracts expose an interface of functions to the network that can be triggered by sending  transactions to the smart contract. Because a smart contract resides on the blockchain, each node can view and execute its instructions, as well as see the log of each interaction with each smart contract. Here, we can view a patient's data created by a wearable device and healthcare providers as two such parties requiring the trust of the blockchain network. Moreover, through other smart contracts, patients can grant and revoke access to their own data as mentioned earlier.

\begin{figure}[h]
    \centering
    \scalebox{.60}{\includegraphics{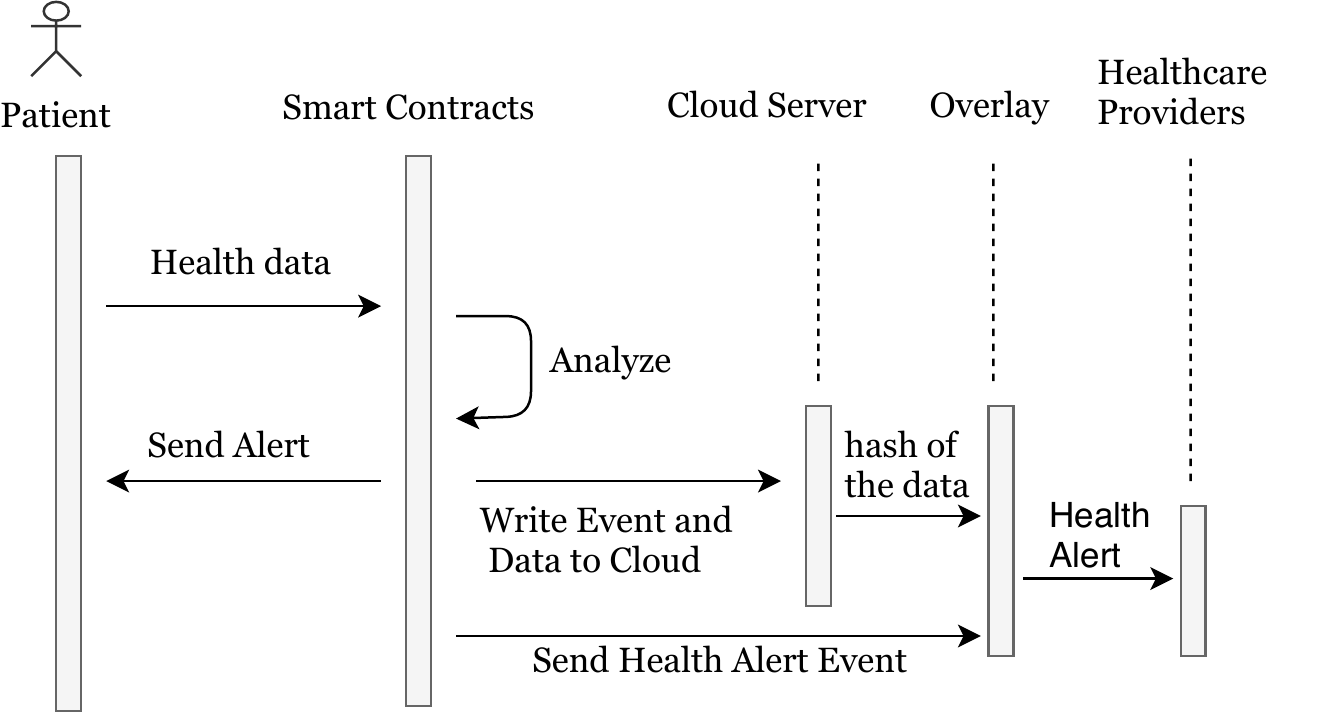}}
    \caption{Logical flow execution of the system} 
      \label{flow}
  \end{figure} 

\subsection{Wearable Devices and Smart Contracts}
A smart contract is executed and sends data to the network when a given condition is met (Figure~\ref{flow}). Consider we set the condition for the highest and lowest level of patient glucose levels in the smart contract itself. If the reading goes up or down beyond this level, the smart contract will issue an alert and send data to the network. To describe the example at hand, lets say \texttt{Patient X} is wearing a device to monitor his/her glucose levels. Constantly, the IoT device will measure \texttt{Patient X}'s glucose levels, and comparing the data to a smart contract, which could potentially be shared between \texttt{Patient X} and \texttt{Hospital Y}. If in the contract \texttt{Hospital Y} wants an alert sent if the glucose levels goes beyond some upper threshold, the event and corresponding data is written to the cloud server, and the hash of the data itself is saved on the Blockchain connected to the overlay network. Finally, a health alert is sent to \texttt{Hospital Y} as indicated as part of the corresponding smart contract.

\subsection*{Cloud Storage:} 
\begin{figure}[h]
    \centering
    \scalebox{.60}{\includegraphics{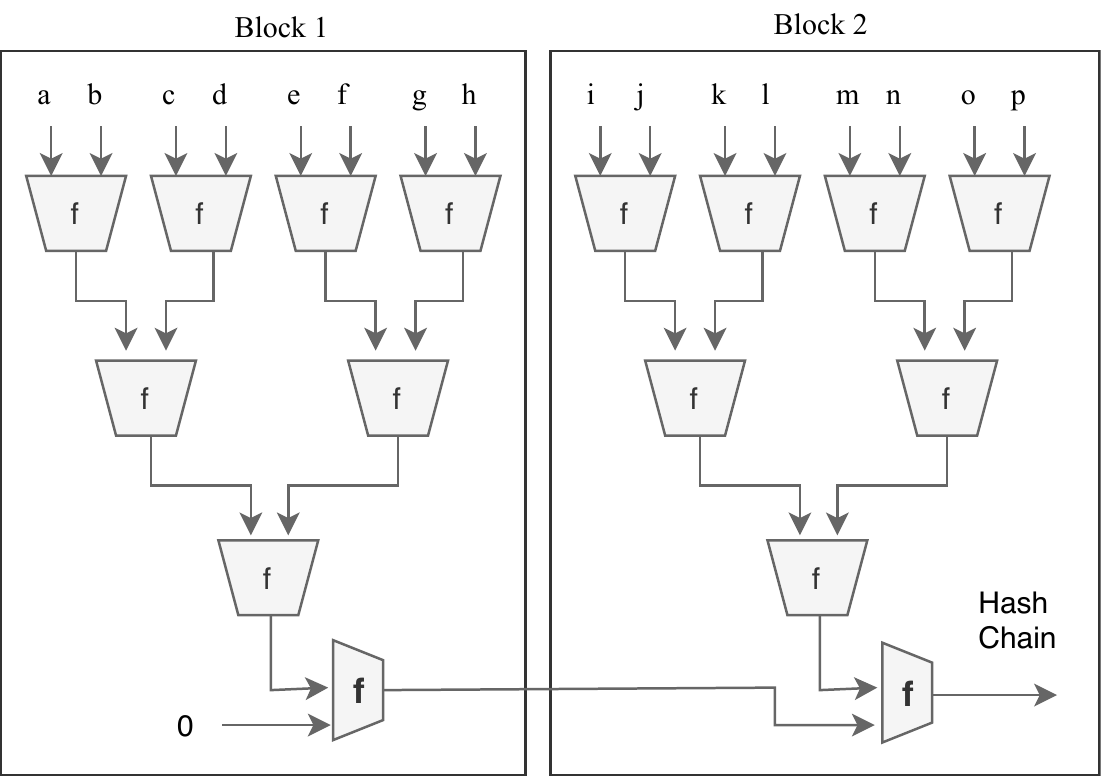}}
    \caption{Merkle Tree} 
      \label{Merkle}
  \end{figure}

We use cloud storage to store the data sent by patient devices. When a smart contract issues an alert for the network it also stores the relevant healthcare data to the cloud. Before sending the data over the cloud, the nodes add a digital signature with the healthcare data. In this instance, we are not storing \textbf{ALL} data that an IoT health device may create, just the data that falls outside of the prescribed \textbf{NORMAL} realm. The cloud servers verify the digital signature added by nodes. In case the digital signature is not correct or not available, data will be discarded by cloud servers. The cloud servers group the user's data in identical blocks and shares the hash of the block to the overlay network. The hash is calculated by a hash function $f$, which could be \texttt{SHA}-$2$ or \texttt{SHA}-$3$. The input could be a long string while the output of the function will be fixed length, e.g. $256$-bit or $512$-bit. A single block contains many documents or strings and we calculate the combined hash of all strings in the form of a Merkle Tree (Figure~\ref{Merkle}). Note that Merkle trees are not a method to store data but is a tree of hashes that makes it easier to verify data efficiently. Also of note is that Merkle trees allow clients to store only the root of the tree (combined hash), not the entire history and therefore original data will be stored in the cloud whereas the combined hash will be transferred to the network.

\subsection{Proof of Authority:}
\texttt{Proof of Authority} (PoA) is a family of consensus algorithms for permissioned blockchain whose prominence is due to performance increases with respect to typical Byzantine Fault Tolerant algorithms; this results from lighter message exchanges. PoA was originally proposed as part of the Ethereum ecosystem for private networks \cite{de2018pbft}. PoA algorithms rely on a set of $N$ trusted nodes called the authorities. Each authority is identified by a unique $ID$ and a majority of them are assumed honest, namely at least $\dfrac{N}{2}+ 1$. The authorities run a consensus to order the transactions issued by clients. Consensus in PoA algorithms relies on a mining rotation schema, a widely used approach to fairly distribute the responsibility of block creation among authorities, thus making it less computationally and more energy efficient than PoW algorithms that were mentioned earlier which have gained prominence in most other cryptocurrency applications. PoA algorithms are much better suited to IoT applications than their counterpart PoW algorithms.

\subsection*{Overlay network:} The overlay network is a peer-to-peer network. The network is based on a distributed architecture. The overlay network is made up of nodes, clusters and each cluster has a Cluster Head. Each node in the network could either be a patient device known as a requestee or a healthcare provider known as a requester. The nodes themselves attached in the network could be mobile devices, computers or tablets. We remove the concept of PoW in our network. We implement \texttt{Proof of Authority} (PoA), instead of PoW as described earlier. We rely on digital signatures of nodes and if a node sends data without a digital signature or with a wrong signature then the node can not add his/her data over the cloud servers and the cloud will not create a hash of such data. To avoid delays (due to low bandwidth) in the network and to reduce overhead, nodes in the network are grouped into clusters. Each cluster elects a Cluster Head ($CH$). We use the election protocol \texttt{SYSTAS} as defined in \cite{kousaridas2015systas}. Each Cluster Head maintains a unique Public Key which is known by all other Cluster Heads. In this manner, when generating new blocks for the chain, the $CH$ can directly authorize the block generator. Each node in a cluster is free to change its cluster and any cluster is free to elect a new $CH$ at any time. The $CH$ of each cluster is responsible from maintaining lists of:

\begin{itemize}
\item Public Keys of Requesters (Healthcare Providers): This allows for $CH$s to maintain a list of entities allowed to access data.
\item Public Keys of Requestees (Patient devices):This allows for $CH$s to maintain a list of entities allowed post data or have their data accessed.
\end{itemize}

The overlay network does not save the healthcare data but only saves the root hash of each block stored in the cloud, in the form of the chain. Due to this property of our model, we eliminate the memory limit problem of blockchain. Each block in the chain contains the hash of the previous block. Every time the cloud sends a hash of the new block, the insertion of the new hash in the chain is called mining. Unlike Bitcoin mining, each Cluster Head decides whether to keep the hash of a new block or not. If the Cluster Head does not want to keep hash of a new block, the hash value of the block is transferred to another cluster head in the network thus limiting overhead.

\section{Transaction Types}\label{trans}
We use both encryption techniques namely, a symmetric algorithm (secret key or session key encryption), which use the same key for both encryptions ($E$) of a plaintext and decryption ($D$) of a ciphertext, whereas in asymmetric algorithms (public key encryption) different keys for encryption of a plaintext and decryption of a ciphertext are used. Generally, data can be encrypted by a public key of any node in the network but data can only be decrypted by the private key of the same node. 

\begin{figure}[h]
    \centering
    \scalebox{.50}{\includegraphics{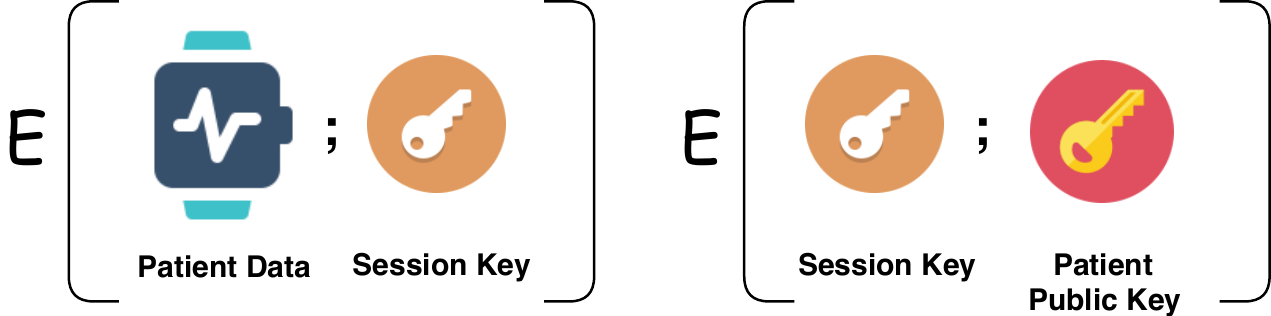}}
    \caption{Transaction 1} 
      \label{Transaction1}
  \end{figure}

We use basically three transaction types in our model. The first transaction (depicted in Figure~\ref{Transaction1}) is the creation of a session key, through which a user/patient can control all access to his/her healthcare data. A patient encrypts his/her data with the help of the session key and stores the data in the cloud and also encrypts the session key with his/her public key. Then, he/she transfers the encrypted session key to the network. The session key can only be decrypted by the patient's private key (as it was encrypted by the patients public key). 

\begin{figure}[h]
    \centering
    \scalebox{.50}{\includegraphics{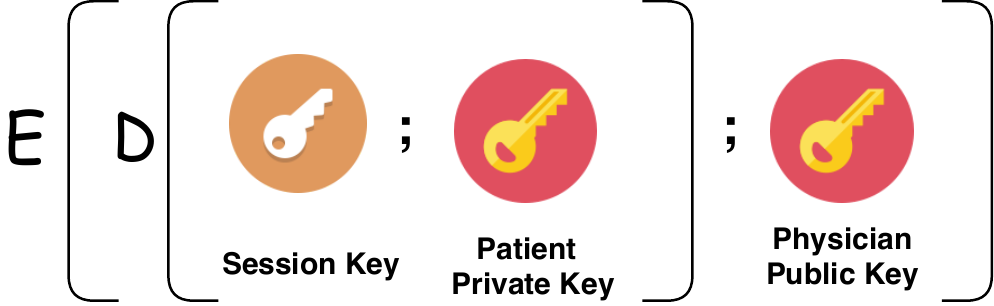}}
    \caption{Transaction 2} 
      \label{Transaction2}
  \end{figure}

Now, the health data can only be accessed by those nodes who have access to the session key of the patient's data, which can easily be stored as part of the smart contracts. If a patient wants to allow a physician or a healthcare provider to access his/her data, the patient will perform a second transaction (see Figure~\ref{Transaction2}). The patient will access his/her session key from the network and then decrypt the key with his/her private key. He/she publishes the session key in the network encrypted with the physician's public key and therefore the physician can decrypt this session key using his/her private key. This process will allow the physician to access medical records of the patient by using the session key. 

\begin{figure}[h]
    \centering
    \scalebox{.50}{\includegraphics{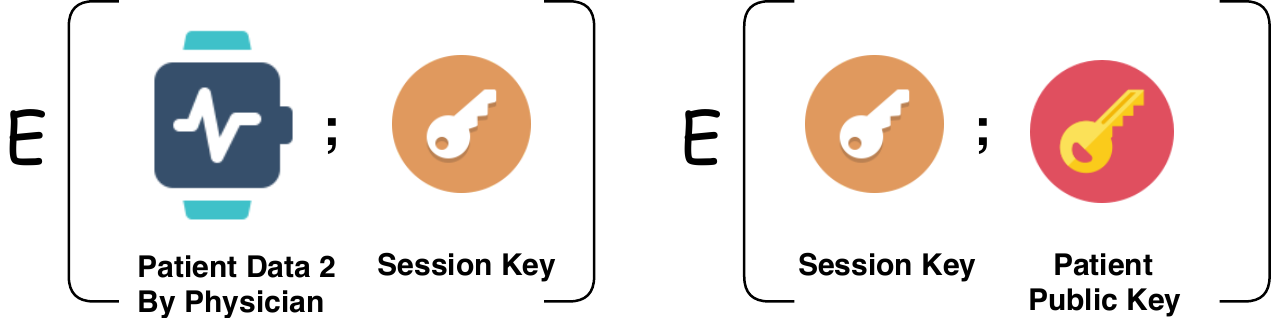}}
    \caption{Transaction 3} 
      \label{Transaction3}
  \end{figure}

On the other hand, the physician may need to send a new medical record to the patient. He can apply the same process of encrypting medical record with a new generated session key and encrypts the session key using the public key of the patient, so that the patient can access the session key with his/her private key, see Fig. \ref{Transaction3}. The physicians can also request to operate patients healthcare devices to the Cluster Head of the network. 

\section{Conclusion}\label{conc}
The incorporation of blockchain technologies in IoT is not an easy task and there are many obstacles in such models. Therefore, the benefits of applying blockchain-based IoT models should be analyzed carefully and taken with caution. Our proposed blockchain-based IoT model removed such obstacles and handles most privacy and security threats while considering the resource-constraints of many IoT devices. This paper has provided the main challenges that IoT and blockchain must address in order to work successfully together. We have also identified the key points where IoT and blockchain can work well together. As a future direction, it would be interesting to see this model tested in a controlled experimental setting testing the performance of the system as a whole. We leave this direction as future work. 

\section*{Acknowledgement}
Research described in this paper was financed by the National Sustainability Program under Grant no. LO1401, and the Ministry of Interior under grant VI20172019093. For the research, the
infrastructure of the SIX Center was used.

\bibliographystyle{IEEEtran}
\bibliography{local_biblio}

\begin{thebibliography}{10}
\providecommand{\url}[1]{#1}
\csname url@samestyle\endcsname
\providecommand{\newblock}{\relax}
\providecommand{\bibinfo}[2]{#2}
\providecommand{\BIBentrySTDinterwordspacing}{\spaceskip=0pt\relax}
\providecommand{\BIBentryALTinterwordstretchfactor}{4}
\providecommand{\BIBentryALTinterwordspacing}{\spaceskip=\fontdimen2\font plus
\BIBentryALTinterwordstretchfactor\fontdimen3\font minus
  \fontdimen4\font\relax}
\providecommand{\BIBforeignlanguage}[2]{{%
\expandafter\ifx\csname l@#1\endcsname\relax
\typeout{** WARNING: IEEEtran.bst: No hyphenation pattern has been}%
\typeout{** loaded for the language `#1'. Using the pattern for}%
\typeout{** the default language instead.}%
\else
\language=\csname l@#1\endcsname
\fi
#2}}
\providecommand{\BIBdecl}{\relax}
\BIBdecl

\bibitem{nakamoto2012bitcoin}
\BIBentryALTinterwordspacing
S.~Nakamoto, ``Bitcoin: A peer-to-peer electronic cash system,'' 2009.
  [Online]. Available: \url{http://www.bitcoin.org/bitcoin.pdf}
\BIBentrySTDinterwordspacing

\bibitem{conoscenti2016}
M.~Conoscenti, A.~Vetro, and J.~C. De~Martin, ``Blockchain for the internet of
  things: A systematic literature review,'' in \emph{Computer Systems and
  Applications (AICCSA), 2016 IEEE/ACS 13th International Conference of}.\hskip
  1em plus 0.5em minus 0.4em\relax IEEE, 2016, pp. 1--6.

\bibitem{kshetri2017}
N.~Kshetri, ``Can blockchain strengthen the internet of things?'' \emph{IT
  Professional}, vol.~19, no.~4, pp. 68--72, 2017.

\bibitem{reyna2018}
A.~Reyna, C.~Mart{\'\i}n, J.~Chen, E.~Soler, and M.~D{\'\i}az, ``On blockchain
  and its integration with iot. challenges and opportunities,'' \emph{Future
  Generation Computer Systems}, 2018.

\bibitem{huh2017}
S.~Huh, S.~Cho, and S.~Kim, ``Managing iot devices using blockchain platform,''
  in \emph{Advanced Communication Technology (ICACT), 2017 19th International
  Conference on}.\hskip 1em plus 0.5em minus 0.4em\relax IEEE, 2017, pp.
  464--467.

\bibitem{dorri2017block}
A.~Dorri, S.~S. Kanhere, R.~Jurdak, and P.~Gauravaram, ``Blockchain for iot
  security and privacy: The case study of a smart home,'' in \emph{Pervasive
  Computing and Communications Workshops (PerCom Workshops), 2017 IEEE
  International Conference on}.\hskip 1em plus 0.5em minus 0.4em\relax IEEE,
  2017, pp. 618--623.

\bibitem{dorri2017tow}
A.~Dorri, S.~S. Kanhere, and R.~Jurdak, ``Towards an optimized blockchain for
  iot,'' in \emph{Proceedings of the Second International Conference on
  Internet-of-Things Design and Implementation}.\hskip 1em plus 0.5em minus
  0.4em\relax ACM, 2017, pp. 173--178.

\bibitem{8624250}
A.~D. {Dwivedi}, P.~{Morawiecki}, and G.~{Srivastava}, ``Differential
  cryptanalysis of round-reduced speck suitable for internet of things
  devices,'' \emph{IEEE Access}, vol.~7, pp. 16\,476--16\,486, 2019.

\bibitem{s19020326}
\BIBentryALTinterwordspacing
A.~D. Dwivedi, G.~Srivastava, S.~Dhar, and R.~Singh, ``A decentralized
  privacy-preserving healthcare blockchain for iot,'' \emph{Sensors}, vol.~19,
  no.~2, 2019. [Online]. Available:
  \url{http://www.mdpi.com/1424-8220/19/2/326}
\BIBentrySTDinterwordspacing

\bibitem{rodrigues2017blockchain}
B.~Rodrigues, T.~Bocek, A.~Lareida, D.~Hausheer, S.~Rafati, and B.~Stiller, ``A
  blockchain-based architecture for collaborative ddos mitigation with smart
  contracts,'' in \emph{IFIP International Conference on Autonomous
  Infrastructure, Management and Security}.\hskip 1em plus 0.5em minus
  0.4em\relax Springer, Cham, 2017, pp. 16--29.

\bibitem{hanada2018smart}
Y.~Hanada, L.~Hsiao, and P.~Levis, ``Smart contracts for machine-to-machine
  communication: Possibilities and limitations,'' in \emph{2018 IEEE
  International Conference on Internet of Things and Intelligence System
  (IOTAIS)}.\hskip 1em plus 0.5em minus 0.4em\relax IEEE, 2018, pp. 130--136.

\bibitem{de2018pbft}
S.~De~Angelis, L.~Aniello, R.~Baldoni, F.~Lombardi, A.~Margheri, and
  V.~Sassone, ``Pbft vs proof-of-authority: applying the cap theorem to
  permissioned blockchain,'' 2018.

\bibitem{kousaridas2015systas}
A.~Kousaridas, S.~Falangitis, P.~Magdalinos, N.~Alonistioti, and M.~Dillinger,
  ``Systas: Density-based algorithm for clusters discovery in wireless
  networks,'' in \emph{2015 IEEE 26th Annual International Symposium on
  Personal, Indoor, and Mobile Radio Communications (PIMRC)}.\hskip 1em plus
  0.5em minus 0.4em\relax IEEE, 2015, pp. 2126--2131.

\end{thebibliography}

\end{document}